\documentstyle[12pt,epsfig]{article}
\newcommand{\slsh}[1]{\not{\hbox{\kern-2pt${#1}$}}}

\newcommand{\ba}[1]{\begin{eqnarray} \label{#1}}
\newcommand{\ea}{\end{eqnarray}}

\def\beq{\begin{equation}}
\def\eeq{\end{equation}}
\def\bea{\begin{eqnarray}}
\def\eea{\end{eqnarray}}

\def\gappeq{\mathrel{\rlap {\raise.5ex\hbox{$>$}}
{\lower.5ex\hbox{$\sim$}}}}

\def\lappeq{\mathrel{\rlap{\raise.5ex\hbox{$<$}}
{\lower.5ex\hbox{$\sim$}}}}

\begin{document}
\pagestyle{empty}


\vspace*{1 cm}
\begin{center}

{\large{\bf Dark Matter and Lepton Flavour Violation in Yukawa Unification with Massive Neutrinos}}

\vspace*{1cm}

M. E. G\'omez, J. Rodr\'{\i}guez-Quintero\\
{\it        Dpto. de Fisica Aplicada, Huelva U., 21071, Huelva, Spain}

S. Lola, P. Naranjo \\
{\it         Department of Physics, University of Patras, 26500 Patras, Greece}

E. Carqu\'{\i}n \\
{\it         Centre of Subatomic Studies, Th. U. Federico Santa Mar\'{\i}a, Valpara\'{\i}so, Chile}

\vspace*{2cm}
{\bf ABSTRACT} \\
\end{center}
\vspace*{5mm}
\noindent
The WMAP dark matter constraints on b-$\tau$ Yukawa Unification 
in the presence of massive neutrinos is revisited. 
The predictions for the bottom quark mass are observed 
to be modified by the large neutrino mixing suggested by the 
data. This enables Yukawa unification also for large $\tan\beta$, and for positive
$\mu$ that were  previously disfavoured. Consequently,
the allowed parameter space for neutralino dark matter differs from the MSSM one.
We also find that the parameter space being compatible with dark
matter also predicts detectable rates for Lepton Flavour Violation at the LHC. 

\vspace{5cm}
{\it Talk given Identification of dark matter 2008, August 18-22, 2008, Stockholm, Sweden}



\newpage

\section{Introduction}\label{intro}

The amount of Cold Dark Matter (CDM) deduced from the Wilkinson Microwave 
Anisotropy Probe (WMAP) data~\cite{Bennett:2003bz,Spergel:2003cb} puts severe 
constraints on possible Dark Matter Candidates, 
inluding the lightest supersymmetric particle (LSP). 
Additional constraints on the model parameters are obtained by
imposing Yukawa unification, and by taking into account the
bounds from Flavour Changing Neutral Currents (FCNC).

In addition to the above, the neutrino data of the past years
provided evidence for the existence of neutrino oscillations and masses,
pointing for the first time to physics beyond the Standard Model.
As expected, the additional interactions required to
generate neutrino masses also affect the energy dependence of the
couplings of the MSSM, and thus modify the Yukawa unification predictions.
A first observation had been that the additional interactions of
neutrinos, which affect the tau mass, may spoil bottom-tau unification 
for small $\tan\beta$ \cite{anant,VB}.
Subsequently, however, it has  been realised that large lepton mixing
naturally restores unification, and even allows Unification for intermediate
values of $\tan\beta$ that were previously disfavoured
\cite{LLR,CELW,GINS}. This is done by making the simple observation
that the $b-\tau$ equality at the GUT scale refers to the
$(3,3)$ entries of the charged lepton and
down quark mass matrices, while the detailed structure of the
mass matrices is not predicted by the Grand Unified
Group itself.
It is then possible to assume mass textures,
such that,{\em  after} diagonalisation at   the
GUT scale, the $(m^{diag}_E)_{33}$ and $(m^{diag}_D)_{33}$
entries are no-longer equal.

The present paper is based on \cite{GLNR,Edson} and 
pay attention to the issues of Dark Matter and Yukawa Unification
taking into account the effects of massive neutrinos and large
lepton mixing in See-Saw models, and 
extending previous results to large $\tan\beta$.
Those issues in mind, we also analyse the prospects for the detection of Lepton Flavour 
violation (LFV) at the LHC via {\it neutralino} decays.


\section{Massive Neutrinos and Unification}

In the presence of massive neutrinos, the predictions for 
$m_b$ and unification clearly get modified. Radiative corrections  
from  the neutrino Yukawa couplings
have to be included in RGE when running from 
$M_{GUT}$ down to $M_{N}$ 
(scale of the heavy right-handed 
neutrinos).  Below $M_N$, right-handed neutrinos decouple from the 
spectrum and an effective  see-saw mechanism is operative; 
the relevant equations are given in \cite{GL}.
In addition, if the GUT scale 
lies significantly below a scale $M_X$, at which gravitational 
effects can no longer be neglected, the 
renormalization of couplings
at scales between $M_X$ and $M_{GUT}$ may induce 
additional effects to the running and the simplest example is
provided by minimal SU(5) \cite{Hisano} (however, 
modifications to soft masses are in this simplest case proportional 
to the $V_{CKM}$ mixing \cite{Hisano}, and thus are significantly suppressed).
Nevertheless, it has been realized  that the influence 
of the runs above the GUT scale on the Dark Matter abundance can be 
very sizeable \cite{Mamb}, due to  changes in
the relation between $m_{\tilde{\tau}}$ and $m_{\chi}$, 
which is crucial in the coannihilation 
area. This we discuss in a subsequent section.

In supersymmetric models, unification 
is  very sensitive to the model parameters, particularly the
Higgs mixing parameter, $\mu$. 
To correctly obtain  pole masses within this framework, 
the Standard model and supersymmetric  threshold corrections
have to be included; for the bottom quark, these
corrections result to a $\Delta m_b$ that 
can be very large, particularly  for large values of $\tan\beta$.
Constraints from BR($b\rightarrow s \gamma$) and
$g_{\mu}-2$ \cite{Bennett:2004pv} are also included in the analysis.
Before passing to the results, however,
let us summarise a few facts on 
the possible range of the mass of the bottom quark:
the 2-$\sigma$ range for 
the  $\overline{MS}$ bottom running mass,
$m_b(m_b)$, is from  4.1-4.4 GeV. Moreover,
$\alpha_s(m_Z)=0.1172 \pm 0.002$, 
and the central  value of $\alpha_s$ corresponds 
to $m_b(m_Z)$ from  2.82 to 3.06 GeV.

\begin{figure}[!h]
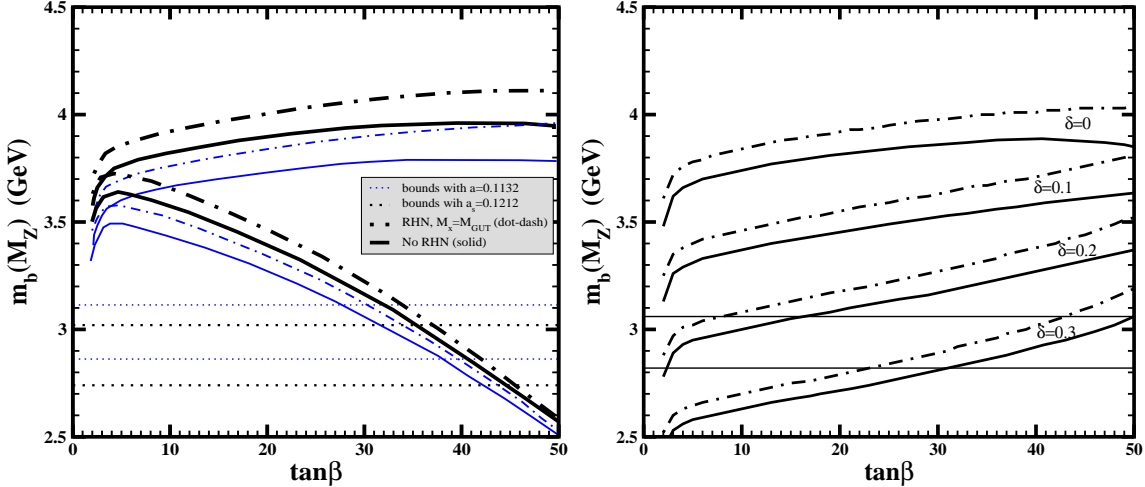

\hspace*{-0.3 cm}
\includegraphics[width=7.5cm,height=6.5cm]{graphs/mbtb_alpha.eps}
\includegraphics[width=7.5cm,height=6.5cm]{graphs/mbtb_xpar.eps}
\caption{\it 
The value of $m_b(M_Z)$ versus $\tan \beta$ assuming
$\alpha_s(M_Z)=0.1212$ (black) and $\alpha_s(M_Z)=0.1132$ (blue), 
$\lambda_b = \lambda_\tau$ at the high scale in the absence 
of lepton mixing for the set of parameters defined in the text. 
The experimental range of $m_b$ (horizontal lines) is also shown, 
for the same values of $\alpha(M_Z)$.
In the right panel, we show $m_b$ as a function of $\tan\beta$, when including
lepton mixing effects for different values of $\delta$ and 
impose $m_b(M_Z)=2.92$ GeVs for $\mu>0$ and $\mu<0$.
}
\label{mb1}
\end{figure}

In Fig. \ref{mb1}, we summarize the predictions for $m_b$
in mSUGRA, where all CP phases are either 
zero or $\pi$ (defining the sign of $\mu$).  In order to discuss 
the dependence of $m_b(M_Z)$ on $\tan\beta$, 
we consider the following representative set of soft parameters: 
$m_{\frac{1}{2}}=800$ GeV, $A_0=0$, $m_0=600$ GeV. 
We study each set for both $\mu>0$ and $\mu<0$; 
setting $\alpha_s(M_z) = 0.1132$ or $\alpha_s(M_z) = 0.1212$ 
(upper and lower experimental bounds).
The figure exhibits the well-known 
fact  that in the absence of phases or
large trilinear terms,  $\Delta m_b$ is positive for $\mu$ positive,
and therefore the theoretical prediction for the $b$ quark pole mass is 
too high to be reconciled with $b-\tau$ 
unification. On the other hand, 
for $\mu<0$,  $\Delta m_b$ is negative and the theoretical prediction 
for the b quark mass can lie within the experimental range 
for values of $\tan\beta$ between roughly 25 and 45;
clearly, for a large $\tan\beta$ it is mandatory
to take into account
the large supersymmetric corrections to $m_b$ 
~\cite{Hal,CW}.  

In Fig. \ref{mb1}, the analysis is performed 
also in the presence of massive neutrinos (dashed lines),
keeping only the third generation couplings , from the $M_{GUT}$ to the
scale of the right handed neutrino masses, 
and evolve the light neutrino
mass operator from this scale down to $M_Z$. 
A large value  of  the Dirac-type 
neutrino Yukawa coupling, $\lambda_N$ at the GUT scale
may arise naturally within the framework of Grand Unification, and
its value is determined by demanding a third generation
low energy neutrino mass of $m_{\nu_3}=0.05$~eV.
The predictions for $m_b(M_Z)$ using the 
lower and upper bounds of the 2-$\sigma$ experimental range of 
$\alpha_s$ and the correponding range for $m_b(M_Z)$ after the evolution of the
bounds on $m_b(m_b)$ are shown for a scale $M_N = 3\times 10^{14}$~GeV.

We observe that for $\mu>0$ the prediction 
for $m_b(M_Z)$ is always very large, despite its dependence on the 
soft terms through $\Delta m_b$.
We have also checked in ref.~\cite{GLNR} that runs above the GUT scale have no 
significant impact (their impact is however not negligeble for our subsequent 
considerations on Dark Matter, as shown in ref.~\cite{Mamb})

The results are significantly modified once we consider the effects of lepton mixing
in the diagonalisation and running of couplings from high to low energies. In order to
show this, we focus on $b-\tau$ 
unification within the framework of SU(5) gauge unification and flavour symmetries that
provide consistent patterns for mass and mixing hierarchies, and naturally
reconcile a small $V_{CKM}$ mixing with a large charged 
lepton one. Taking into account the particle content of SU(5) representations
(with symmetric up-type mass matrices, and down-type mass matrices that are transpose to
the ones for charged leptons), 
one finds the approximate relations
\begin{equation}
{\mathcal {M}}_d^0 \propto A\,\left(\begin{array}{cc}
0 & 0 \\
x & 1  
\end{array}\right), \,\,\,\,
{\mathcal {M}}_{\ell}^0 \propto A\,\left(\begin{array}{cc}
0 & x \\
0 & 1 
\end{array}\right)
\label{Ml}
\end{equation}
which, after diagonalization, lead to 
\begin{equation}
\frac{m_b^0}{1+x^2}=\frac{m_{\tau}^0}{1-x^2}\,\rightarrow \,
m_b^0=m_{\tau}^0\,\left(1-\underbrace{2x^2}_{\delta}+{\mathcal{O}}
\left(\delta ^2\right)\right)
\label{btau}
\end{equation}
where $\delta$ parametrises  the flavour mixing in the (2,3) sector. 

In the right pannel of Fig.\ref{mb1} we show the change of $m_b$ as a 
function of $\tan\beta$, when the effects from large lepton mixing are appropriately
considered. Comparing with the previous plots,  we 
see how solutions with positive
$\mu$ are now viable, for the whole range of $\tan\beta$.
The appropriate size of the parameter $\delta$ in each case 
can be determined by imposing 
the relation $\lambda_\tau=\lambda_b(1+\delta)$ at $M_{GUT} $ and investigating
the values that are required in order to obtain a correct prediction for  $m_b(M_Z)$.


\section{Dark Matter constraints and Yukawa unification}

In mSUGRA (or the CMSSM) for choices of soft terms below the 
TeV scale, the LSP is Bino like and the prediction for $\Omega_\chi h^2$ is 
typically too large for models that satisfy the experimental constraints on 
SUSY. In fact, the values of WMAP can essentially  be obtained in two regions:
(i) $\chi-\tilde{\tau}$ coannihilation region that occurs for 
$m_\chi\sim m_{\tilde{\tau}}$; and
(ii) that of resonances in the $\chi-\chi$ annihilation channel, which occur 
for  $m_A\sim 2 m_\chi$.

Since the above areas are ``fine-tuned'',  they will inevitably be 
sensitive to the changes induced by GUT unification and sizeable mixing in
the charged lepton sector. 
The runs corresponding to 
$M_X>M_{GUT}$ have a big impact on the 
neutralino relic density. The large
values of the gauge unified coupling $\alpha_{SU(5)}$ tend to 
increase the values
of   $m_{\tilde{\tau}}$, even if we start with small $m_0$ at $M_X$.

\begin{figure}[!h]
\begin{center}
\begin{tabular}{cc}
\includegraphics[width=6.5cm,height=5.5cm]{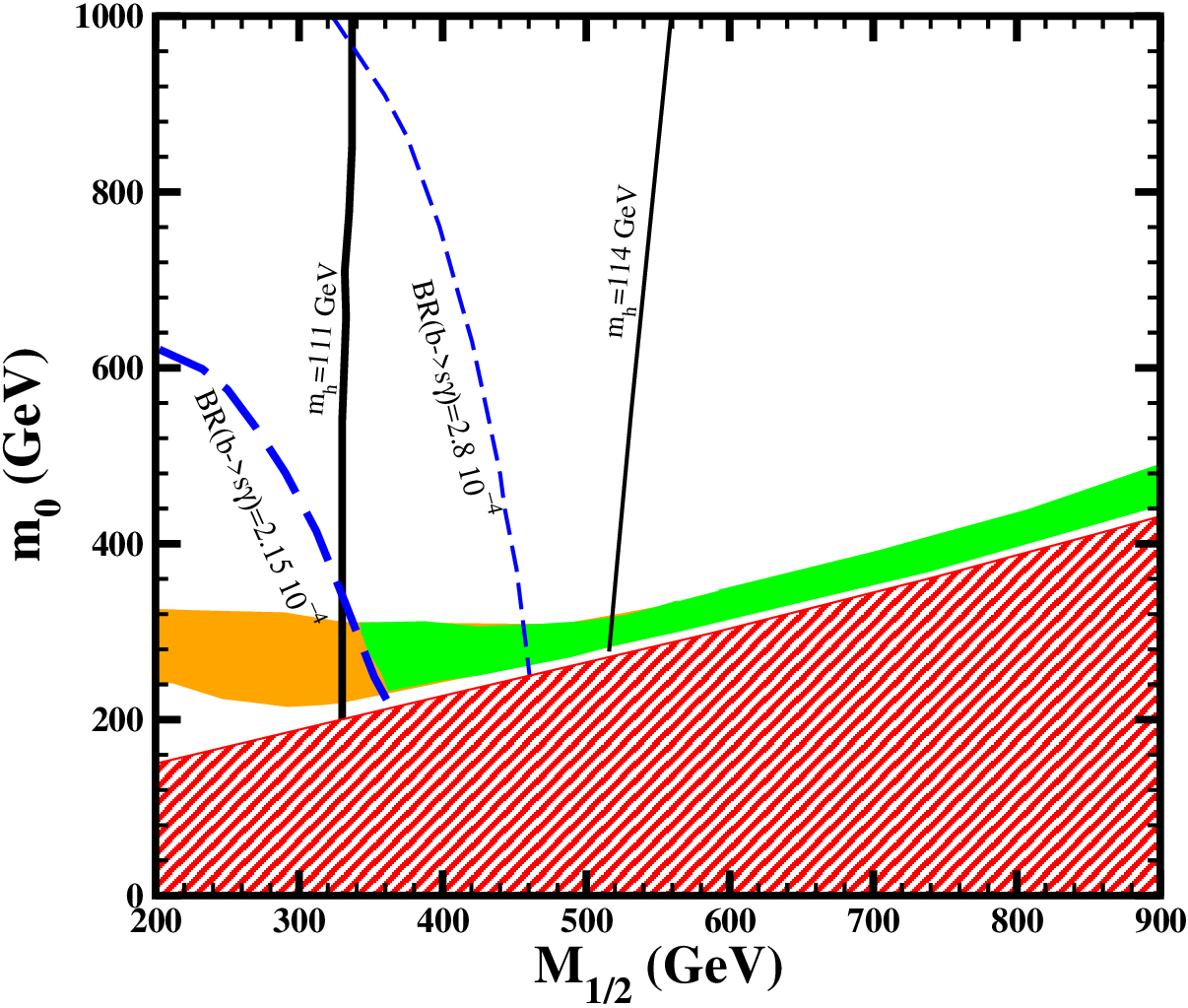} &
\includegraphics[width=6.5cm,height=5.5cm]{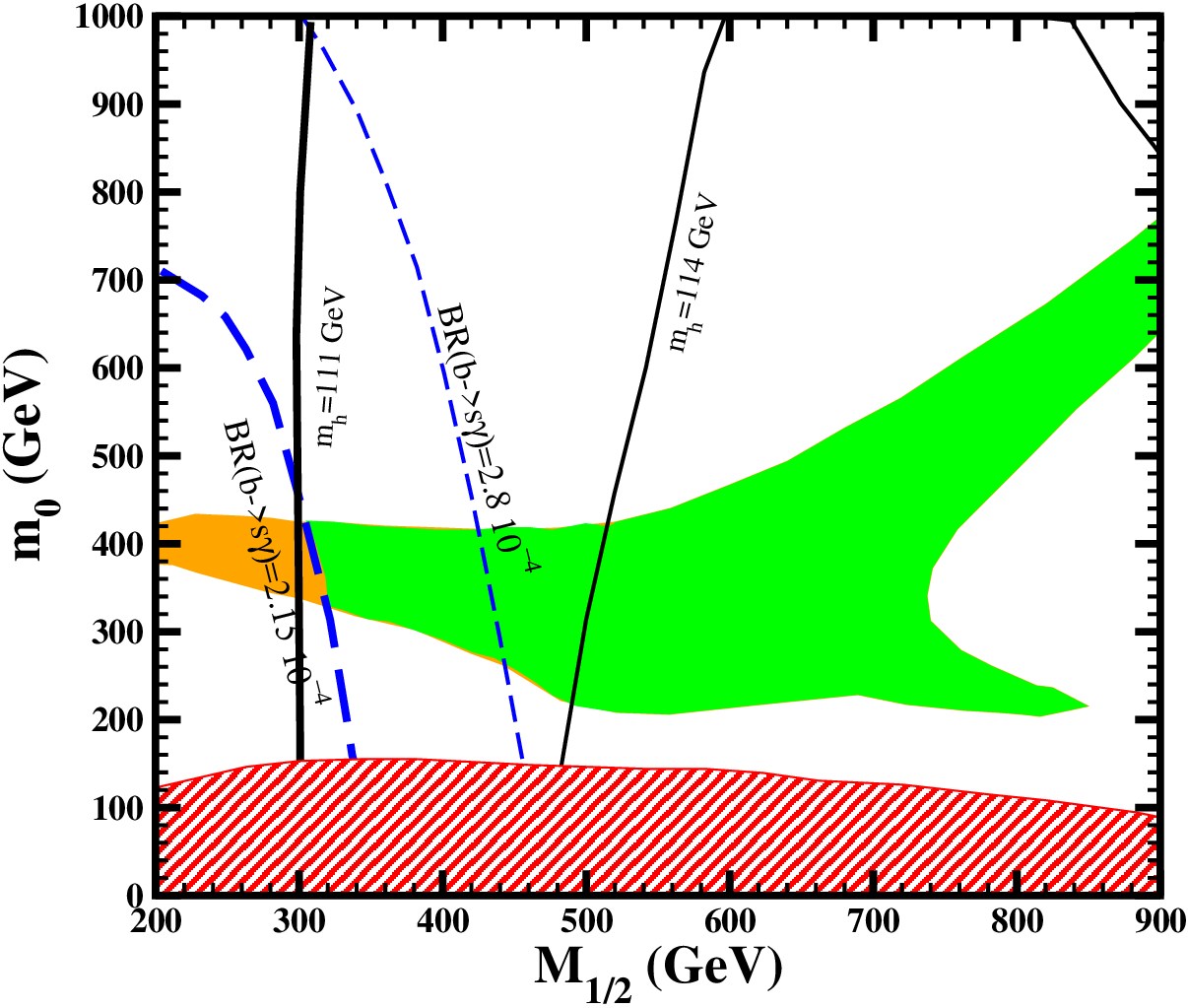} \\
\includegraphics[width=6.5cm,height=5.5cm]{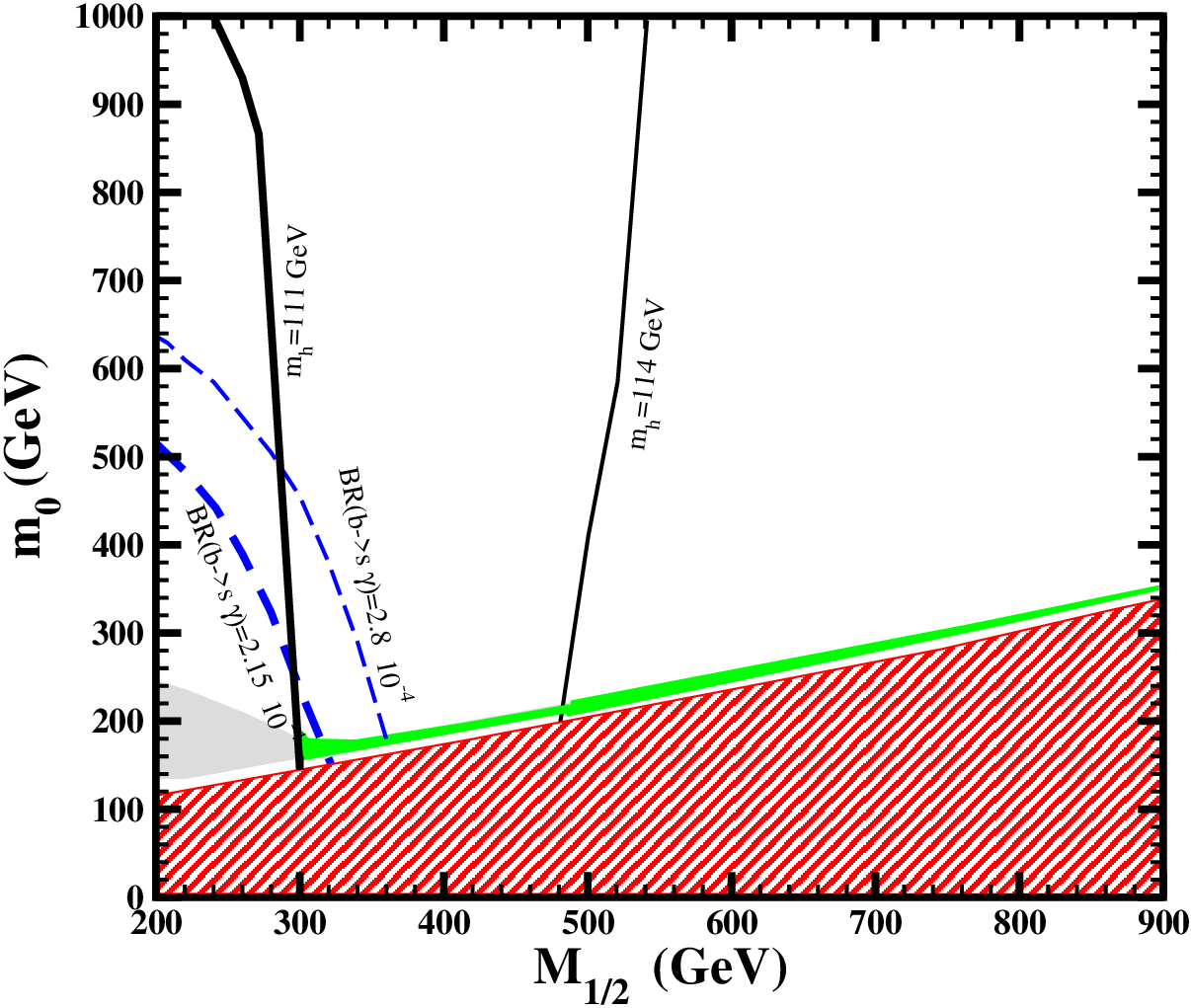} &
\includegraphics[width=6.5cm,height=5.5cm]{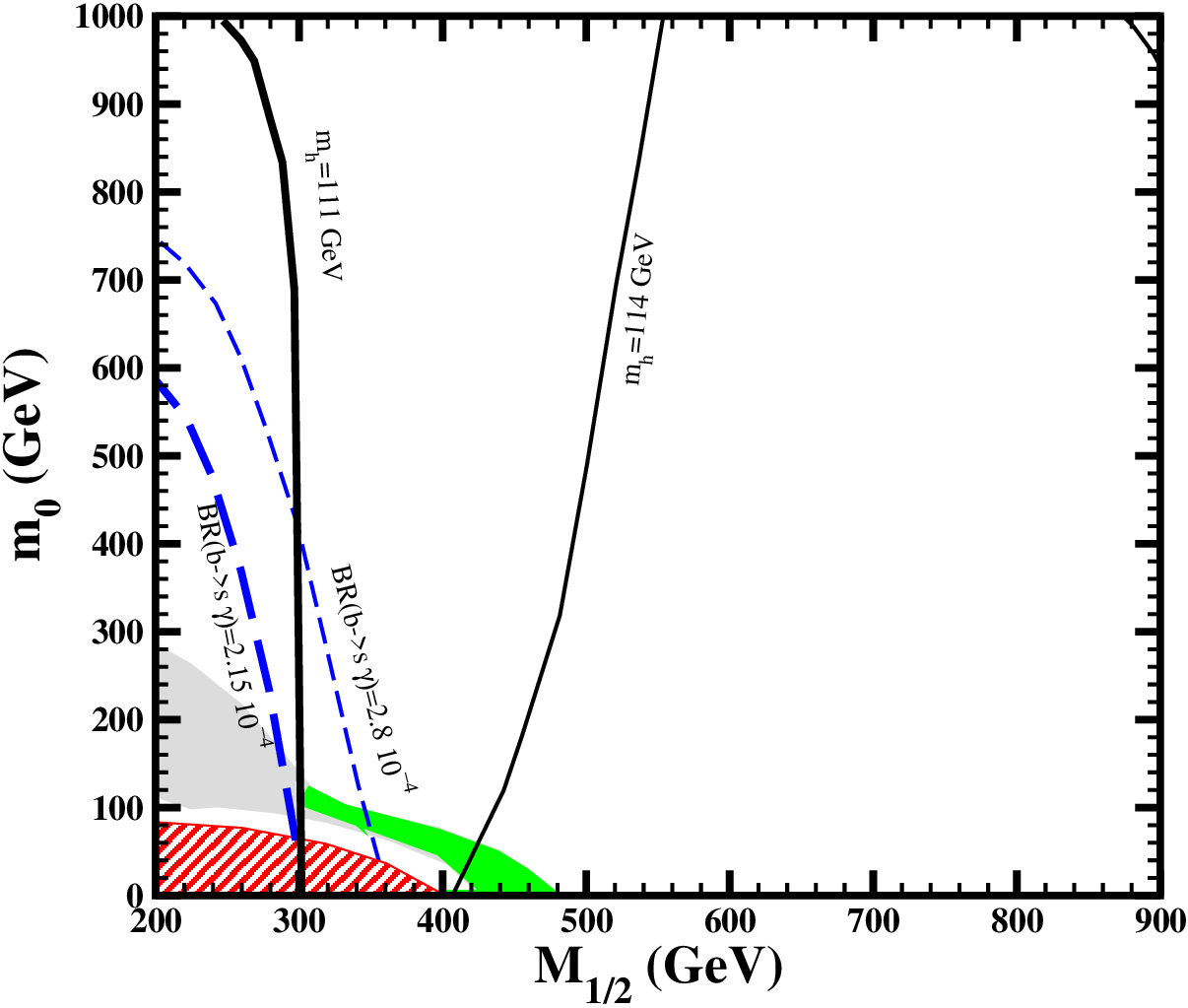}
\end{tabular}
\end{center}
\caption{\it WMAP area for the case of $\tan \beta=45$ (top) and 
$\tan\beta=35$ (down), $\mu>0$, $A_0=m_0$,
the central value of the bottom mass $m_b\left(M_Z\right)=2.92$ GeV and 
$\delta \sim 0.42$ without (left) and with (right) the $SU(5)$ running.}
\label{area45}
\end{figure}


We see that the consideration of mixing effects, in combination to
the inclusion of effects from the runs above $M_{GUT}$, significantly enhances 
the allowed parameter space (green area), an effect that is more visible
for large $\tan\beta$. The reduction of the allowed parameter space for smaller 
values of $\tan\beta$, is already evident for $\tan\beta =35$. 
The case with $\mu<0$ and a more detailed discussion of lepton mixing
effects are presented in \cite{GLNR}.

\begin{figure}[!h]
\begin{center}
\begin{tabular}{cc}
\hspace*{-0.3cm}
\includegraphics[width=6.5cm,height=5.5cm]{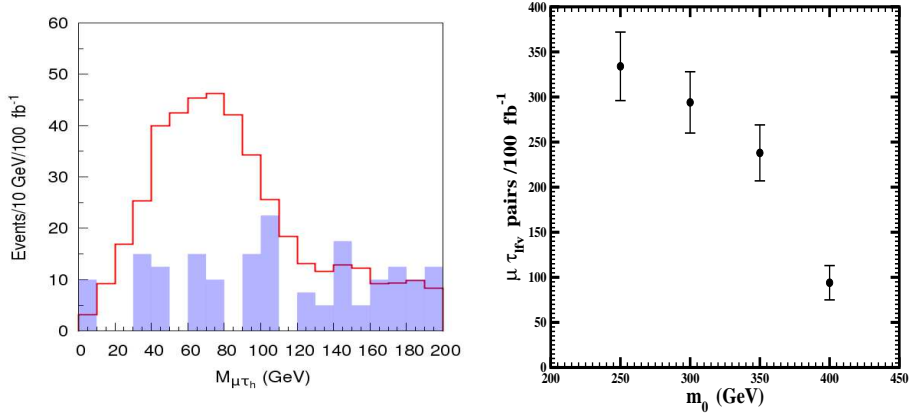}  &
\includegraphics[width=5.5cm,height=5.5cm]{graphs/eventnumber.eps}
\end{tabular}
\end{center}
\caption{\small \it  Left: the signal for  excess $\mu\tau_h$ LFV pairs (red solid lines) 
and subtracted $\mu\tau_h - e\tau_h$ Standard Model backgrounds (shaded) for 
an optimal set of parameters allowed by WMAP constraints. 
Right: number of LFV pairs for different values of $m_0$ and $M_{1/2}$ fixed at 500 GeV. 
(see \cite{Edson} for the complete definition of parameters).} 
\label{plot-edson}
\end{figure}

In ref.~\cite{Edson}, we study the optimal conditions within the above-described 
theoretical framework for $\tau$ flavour violation to be observable 
in $\chi_2 \to \chi + \tau^{\pm} \mu^{\mp}$ at LHC. 
Fig.~\ref{plot-edson}, taken from~\cite{Edson}, 
shows the simulated signal (using PYTHIA) for LHC conditions 
of lepton number violation, as the excess of $\mu-\tau$ over $e-\tau$ pairs, 
for an optimal set of parameters allowed by the WMAP constraints 
in Fig.~\ref{area45}.

\section{Conclusions}

We revisited the WMAP dark matter constraints on Yukawa Unification 
in the presence of massive neutrinos. Large neutrino mixing, as indicated 
by the data modifies the predictions for the bottom quark mass,
and enables Yukawa  also for large $\tan\beta$ and for positive
$\mu$ that were  previously disfavoured.
A direct outcome is that the allowed parameter space
for neutralino  dark matter also increases, particularly 
for areas that are tuned, namely the ones with resonant enhancement of the 
neutralino relic density.

For completeness, we also note that for the cosmologically favoured parameter
region, we found lepton flavour violating rates very close to the current
experimental bounds \cite{GLNR} and study its possible detection at 
LHC. 
Finally, interesting effects may arise in the case of non-universal
soft terms. These are also discussed in detail in \cite{GLNR}.

\vspace*{0.5cm}

{\bf Acknowledgements:} 
The research of S.L. and P.N. is funded by the FP6 
Marie Curie Excellence Grant MEXT-CT-2004-014297. That of 
M.E.G. and J.R-Q. is supported by spanish MEC project FPA-2006-13825
and P07FQM02962 of Junta de Andaluc\'{\i}a. The work of E.C. 
has been partly supported by MECESUP Chile grant and HELEN program.



\end{document}